\documentclass{www13-companion-accepted}

\usepackage{graphicx}
\usepackage{acronym}
\usepackage{subfigure}
\usepackage{booktabs}
\usepackage{algorithm}
\usepackage{algcompatible}
\usepackage{url}
\usepackage{amstext}
\usepackage[utf8]{inputenc}
\usepackage{amsmath,amssymb}

\usepackage{algorithm}
\usepackage{algpseudocode}

\acrodef{P2P}{Peer-to-Peer}
\acrodef{TTL}{Time-To-Live}

\begin{document}
%


\title{Resilience of Dynamic Overlays through Local Interactions}

\numberofauthors{1}
\author{
\alignauthor
Stefano Ferretti\\
       \affaddr{Department of Computer Science and Engineering}\\
       \affaddr{University of Bologna, Italy}\\
       \affaddr{Bologna, Italy}\\
       \email{s.ferretti@unibo.it}
}

\maketitle              

\footnotetext{Please cite as: \textbf{Proceedings of  5th International Workshop on Simplifying Complex Networks for Pratictioners (SIMPLEX 2013) - World Wide Web Conference (WWW 2013), ACM, Rio de Janeiro (Brazil), May 2013}.}

\begin{abstract}
This paper presents a self-organizing protocol for dynamic (unstructured P2P) overlay networks, which allows to react to the variability of node arrivals and departures.
Through local interactions, the protocol avoids that the departure of nodes causes a partitioning of the overlay. We show that it is sufficient to have knowledge about 1st and 2nd neighbours, plus a simple interaction P2P protocol, to make unstructured networks resilient to node faults.
A simulation assessment over different kinds of overlay networks demonstrates the viability of the proposal.
\end{abstract}


\section{Introduction}

\ac{P2P} networks represent a specific use case that can be analyzed through complex network theory \cite{gridpeer,simplex}.
Fault tolerance of \ac{P2P} networks is extremely important and a plethora of studies examines the resilience dynamics in case of churn, i.e.~high rate of node arrivals and departures \cite{Loguinov:2003,Luciano1}. The main motivation is to understand whether algorithms locally executed by peers guarantee network resilience and allow preserving the characteristics of the P2P overlay, such as network connectivity and routing performance. 
Actually, most of these studies focus on structured P2P networks \cite{Kong:2008,Wang:2006}.
Structured \ac{P2P} overlay networks are those where links among nodes are created based on the contents hold by nodes. Examples of structured architectures are tree-based and hierarchical content-dependent structures, as well as \ac{P2P} systems built using Distributed Hash Tables (DHTs) \cite{Basu05nodewiz,Cai03maan,Hidalgo:2011}. 
Due to the need of maintaining such structure, reconfiguration algorithms are usually executed to preserve the general characteristics of the net.

Conversely, in an unstructured \ac{P2P}~overlay network, links among nodes are established arbitrarily, i.e.~they do not depend on the contents being disseminated \cite{EberspacherS05a,simplex,Leitao}.
These solutions are particularly simple to build. Thus, unstructured overlays may be useful in very dynamic contexts. 
Due to this ``freedom'' in the creation and management of the overlay, an option is to avoid the definition of protocols in charge of reacting to node departures, leaving the overlay management to the attachment process only \cite{shimada}. This commonly guarantees that certain network properties are preserved on a steady state \cite{gridpeer}.
However, in case of multiple node departures, some of the properties of these overlays may disappear \cite{Leonard:2005}. For instance, the overlay might get partitioned upon failure of links connecting different clusters. 
Alternatively, some important links might be lost that were playing a main role to keep a limited network diameter; as an example, in small worlds there are links among distant nodes, that strongly reduce the average shortest path length.
While the P2P network is unstructured, it has a topology that provides certain characteristics. These should be maintained, at least up to a certain extent, in order to provide some guarantees and the ability of the network to spread contents.

The aim of this work is to study a local mechanism, which is in charge of reacting to node churn without the need to introduce a structure on the P2P overlay. 
We are looking for a decentralized algorithm that enables nodes to react to network changes, so that the general characteristics of the network topology are preserved.
The mechanism requires that each node $n$ has knowledge of its 1st neighbours (i.e.~friend nodes, directly connected to $n$) and 2nd neighbours (i.e.~friends of $n$'s friends) only. Upon a neighbour departure, each node is able to understand if some 2nd neighbour is no more reachable. 
If this is the case, the node creates a link with it; a very simple communication protocol is employed to coordinate peers and avoid that multiple neighbours create a novel link with the same node. This way, no multiple links are created that might alter the unstructured network topology.

The approach is thought to preserve links that connect different components of the network. It is usual to have some central node that routes an important part of nodes' messages. In complex network theory, several measures have been introduced to characterize this phenomenon, e.g.~betweenness centrality \cite{Bader:2007,Newman200539}.
The calculation of these metrics usually involves a full (or partially full) knowledge about the network. Conversely, the aim of this work is to preserve connectivity without such a global knowledge, despite the failure of some central node $n$, by replacing failed links with novel links among neighbours of $n$ (which were not neighbours before the $n$ failure). Some previous proposals dealt with similar problems, e.g~\cite{massoulie,VoulgarisGS05}.

It is thus interesting to observe how the protocol performs over networks with different connected clusters. 
A simulation assessment is presented that studies the protocol over uniform networks, where links are created by randomly choosing nodes as neighbours, and clustered networks. Results demonstrate that the presented approach preserves network connectivity and resilience, despite node churn.

%

The remainder of this paper is organized as follows. Section \ref{sec:prot} presents the P2P protocol. Section \ref{sec:eval} describes the simulation environment and discusses the obtained results. Finally, Section \ref{sec:conc} provides some concluding remarks.

\section{The Protocol}\label{sec:prot}

Every node $n$ maintains the list of its neigh\-bours (1st neighbours, $\Pi_n$), and the neighbours of its neighbours (2nd neigh\-bours, $\Pi^2_n$).
The degree of $n$ is the amount of 1st neighbours, i.e.~$|\Pi_n|$. Every time the list of 1st neighbours $\Pi_n$
changes, due to some node arrival or departure, $n$ informs its other 1st neighbours of this update.
With $\Pi^2_{n|m} = \Pi_m - \Pi_n$, we identify the $n$'s 2nd neighbours which can be reached through $m$. Hence, $\Pi^2_n = \cup_{k \in \Pi_n} \Pi^2_{n|k}$.

\begin{figure}[thbp]
   \centering
   \includegraphics[width=\linewidth]{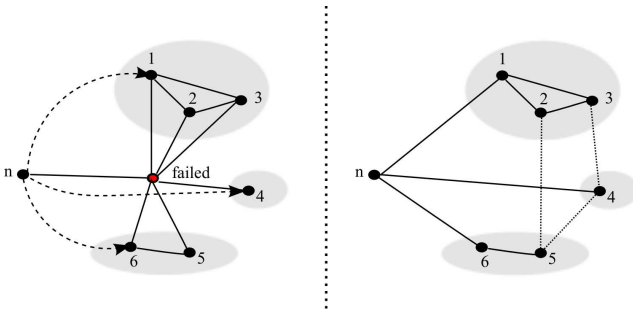}
   \caption{Reconfiguration at $n$ after a node failure}
   \label{fig:schema}
\end{figure}

The aim of the protocol is to avoid that a node failure causes a network partitioning and does not increase excessively the distance among nodes in terms of hops. 
It is well known that in networks, certain links can play a main role to keep a limited network diameter. For instance, small worlds are characterized by links among distant nodes, that strongly reduce the average shortest path length.
With this in view, when a node $f$ fails, each neighbour $n \in \Pi_f$, that was employing $f$ to reach some of its 2nd neighbours $\Pi^2_{n|f}$, reconfigures its links so that these nodes still remain in its 2nd neighbourhood, i.e.~$n$ checks that nodes in $\Pi^2_{n|f}$ remain in its 2nd neighbourhood or it creates links with them. 
Figure \ref{fig:schema} shows a related example, that focuses on links that should be created at $n$, after a neighbour failure, so that its 2nd neighborhood remains unaltered. Upon failure of the central node, novel links are created between $n$ and a node of each cluster. This has the two following advantages: i) the network does not get disconnected, and ii) a minimum number of links is created. (Note that the example focuses on $n$; however, other links might be necessary for other nodes.)

According to the scheme, when a node $f \in \Pi_n$ fails, $\forall p \in \Pi_f$ there are three possible cases:\\
\begin{itemize}
 \item $p \in \Pi_n$ : $n$ and $p$ are neighbours; there is nothing to do;
 \item $p \notin \Pi_n$, but $p \in \Pi^2_n$ since $p \in \Pi^2_{n|q}$ for some $q \in \Pi_n, q \neq f$: $p$ is still a 2nd neighbour of $n$; there is nothing to do;
 \item $p \notin \Pi_n, p \notin \Pi^2_n$ : $p$ is no more a 2nd neighbour of $n$; $n$ takes part to the distributed procedure to create a link with $p$ (explained below).
\end{itemize}
In essence, links are created among nodes in different clusters which were connected through $f$ only.



The distributed protocol is reported in Algorithms \ref{alg:active}-\ref{alg:passive} and works as follows. As mentioned, the procedure is activated at node $n$, after the failure of one of its neighbours, $f \in \Pi_n$. All nodes $p \in \Pi_f \wedge p \notin \Pi_n \wedge p \notin \Pi^2_n$ are considered. 
The protocol is made of two distinct parts: active (Algorithm \ref{alg:active}) and passive (Algorithm \ref{alg:passive}) behavior.
The node $n$ executes the active behavior to create a link with some node only while its actual degree does not surpass a given threshold value (Algorithm \ref{alg:active}, line \ref{code:control}). This is a practical control to avoid that the degree of a node grows out of control and that the network topology changes radically. (Note that the threshold must not be too low, otherwise it would contrast the creation of additional links, and this might generate network partitions.)
Then, $n$ waits a random time (line \ref{code:wait}). This is a typical contention-based approach, used to diminish the probability that multiple nodes of the same cluster attempt to create a novel link with $p$ at the same time \cite{PalazziFR09}. Then, $n$ randomly selects a node $p$ from the list of nodes to which the cluster needs to be linked, and sends it a link creation request (lines \ref{code:extract}--\ref{code:req}). These steps are repeated until the list of lost 2nd neighbours becomes empty (line \ref{code:control}).

\begin{algorithm}[htbp]
\caption{Failure management of $f$ at node $n$: Active behavior}
\label{alg:active}
\begin{algorithmic}[1]
\State $P \gets \{ p \in \Pi_f | \ p \notin \Pi_n, p \notin \Pi^2_n \}$
\Statex
\While {($P \neq \emptyset) \wedge (| \Pi_n | \leq \text{thresholdDegree})$}\label{code:control}
  \State wait random time \label{code:wait}
  \State $p \gets$ extract random node from $P$ \label{code:extract}
  \State send link creation request to $p$\label{code:req}
\EndWhile
\end{algorithmic}
\end{algorithm}

The node reacts to received messages as follows (Algorithm \ref{alg:passive}). Upon a message reception, answering a previous link creation request from $n$ to $p$ (lines \ref{code:ans_b}--\ref{code:ans_e}), if the answer is positive, then $n$ informs all its neighbours about its novel neighbour and creates the link with $p$. 
Upon reception at $n$ of a message from a neighbour $q$, about a novel link creation between $q$ and a node $m$ (line \ref{code:l_creation}), $n$ removes $m$ from $P$, i.e.~the list of nodes that the cluster needs to be linked with (line \ref{code:P}). Moreover, $n$ adds in its cache information about $m$, so that $n$ knows in the future that $m$ is a 2nd neighbour, reachable through $q$, i.e.~$m \in \Pi^2_{n|q}$ (line \ref{code:2nd_n}). 
Upon reception of a message from a node $p$ asking $n$ to become neighbours, $n$ accepts the request only if $p$ is not a 1st or 2nd neighbour of $n$ (it is possible that some of its neighbours just created a link with $p$, lines \ref{code:l_creat_req_b}--\ref{code:l_creat_req_e}). In this case, it sends a positive answer to $p$ and informs all the $n$ neighbours of this novel link $(n, p)$; finally, it adds $p$ to its neighbours.

\begin{algorithm}[htbp]
\caption{Failure management of $f$ at node $n$: Passive behavior}
\label{alg:passive}
\begin{algorithmic}[1]
\State $P \gets \{ p \in \Pi_f | \ p \notin \Pi_n, p \notin \Pi^2_n \}$\label{code:P}
\Statex 
\Require message from $p$ answering a link creation request
  \If {answer is OK}\label{code:ans_b}
    \ForAll{$m \in \Pi_n$}
      \State send message to $m$ for link creation $(n, p)$
    \EndFor
    \State add $p$ to $\Pi_n$
  \EndIf\label{code:ans_e}
\Statex 
\Require message from $q \in \Pi_n$, for link creation $(q, m), m \in P$
  \State extract $m$ from $P$\label{code:l_creation}
  \State add $m$ to $\Pi^2_{n|q}$\label{code:2nd_n}
\Statex 
\Require message from $p$ with a link creation request
  \If{$p \in \Pi^2_n$}\label{code:l_creat_req_b}
    \State send refuse message
  \Else
    \State send accept message
    \ForAll{$m \in \Pi_n$}
      \State send message to $m$ for link creation $(n, p)$
    \EndFor
    \State add $p$ to $\Pi_n$
  \EndIf\label{code:l_creat_req_e}
\end{algorithmic}
\end{algorithm}


The protocol avoids that multiple link creation requests are generated from nodes of the same cluster towards a given lost 2nd neighbour. This reduces the amount of messages sent in the network and avoids that the network topology is altered significantly. 

\section{Performance Evaluation}\label{sec:eval}

To assess the proposed protocol, we simulated the protocol over different network topologies. In the following, we describe the simulation settings and the obtained results when running the protocol over uniform networks and clustered networks. 
Actually, during our experimental evaluation we employed also random graphs, obtaining results comparable to those obtained for uniform networks (hence, for the sake of brevity we do not show them in the paper). 

\subsection{Simulation}

The simulator was written in GNU Octave.
As mentioned, two types of unstructured networks were considered: uniform networks and clustered networks. 
Uniform networks are those where all nodes start with the same degree. Then, due to node failures and arrivals (and the reconfiguration imposed by the P2P protocol), the node degree might change.
We varied the initial degree of nodes. During our tests, we set the threshold value for creating novel links in the protocol (i.e.~thresholdDegree in the algorithm) equal to the initial degree, i.e.~the node runs the active behavior when its actual degree is below the initial degree.

As a matter of fact, the P2P protocol is thought for those networks that have important links that connect different parts of the network; thus, it is interesting to observe how the protocol performs over nets composed of different connected clusters.
In these simulations, network clusters were set to be of the same size.
We set two different parameters to create the network. The first parameter is the probability $\gamma$ of creating a link among nodes of the same cluster. Each node is linked to another node of the same cluster with a probability $\gamma$; hence, inside a cluster, nodes are organized as a classic random graph.
As to inter-cluster links, the amount of links created between the two clusters was determined based on a certain probability $\omega$ times the number of nodes in the clusters (i.e.~each node has a probability $\omega$ of having a link with each external cluster).

Upon a node failure, all its links with other nodes are removed. The failed node is randomly selected among those active at that simulation step. Then, the node passes to an inactive state; it can be selected further on to simulate a novel node arrival.
Thus, a node arrival is simulated by changing the state of a randomly selected inactive node to pass to the active state. This event triggers the creation of novel links with other randomly selected nodes. 
Different joining procedures were executed, depending on the network topology under investigation. 
As concerns uniform networks, a random set of neighbours was selected, whose size was equal to the fixed degree that characterized the starting network topology.
As concerns clustered networks, the node was associated to a cluster, and links with nodes in that cluster were randomly created based on the $\gamma$ probability, as in a classic random graph. Then, for each other cluster, the node creates, with probability $\omega$, a link with a random node of that cluster.

%
%

\subsection{Evolution with a stable network size}

\begin{figure}[thbp]
   \centering
   \includegraphics[width=\linewidth]{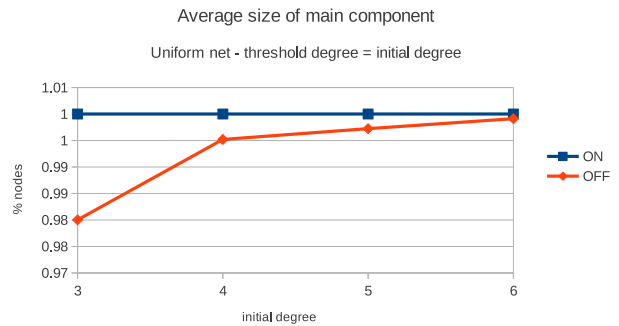}
   \caption{Main component size -- uniform network}
   \label{fig:evoluzione_unif_comp}
\end{figure}

\begin{figure}[thbp]
   \centering
   \includegraphics[width=\linewidth]{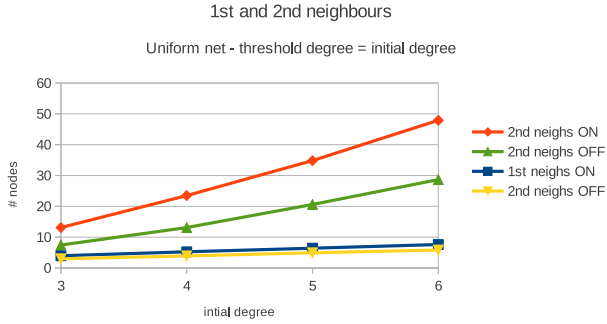}
   \caption{Average amount of 1st and 2nd neighbours -- uniform network}
   \label{fig:evoluzione_unif_neighs}
\end{figure}

In this section we study the performance of the P2P protocol when the network is almost stable, i.e.~when the node failure rate is equal to the node arrival rate. The network evolves with nodes leaving and joining the network, but the net size does not change during the system evolution.
We present averages of obtained results from a corpus of $20$ simulations for the same scenario. During the simulations we removed the transient from the analyzed logs.

First, we consider the case of uniform networks. During the simulations we varied several parameters such as network size and initial fixed node degree. Obtained results were similar in all the simulations, leading always to the same conclusions (hence, we show a limited number of scenarios). 

Figure \ref{fig:evoluzione_unif_comp} shows the average fraction of nodes that are in the main component when the P2P protocol is executed (ON mode) or not (OFF mode).
In this case, a network of $200$ nodes is considered, and the initial fixed degree was varied. 
As shown in the chart, the majority of nodes remain in the main component during the network evolution. This is perfectly reasonable since only a single node failure and node arrival was simulated at each simulation step. Hence, a uniform network cannot be subjected to important network partitions in this particular scenario. However, it is interesting to observe that while some nodes exit from the main component in the OFF mode, all the active nodes remain in the main component in the ON mode.

Figure \ref{fig:evoluzione_unif_neighs} shows the average number of 1st and 2nd neighbours of each node. It is possible to notice that, as expected, the higher the initial degree the higher the number of (1st and 2nd) neighbours. Moreover, the ON mode provides a higher amount of neighbours w.r.t.~OFF mode. This corresponds to a higher connectivity for network nodes.

When we consider a clustered network, we expect a similar behavior, with the exception that if nodes fail that have links which connect different clusters, then the network might disconnect. 
In these simulations, a clustered network was randomly created (based on $\gamma, \omega$); then, a certain amount of nodes was randomly selected, and these nodes were forced to fail (in the following we show results when $5$ nodes were forced to fail at the beginning of the simulation). Then, at each simulation step a single node was randomly selected to fail and a novel node arrival was triggered.

\begin{figure*}[thbp]
   \centering
   \subfigure[step $0$]{\includegraphics[width=0.48\linewidth]{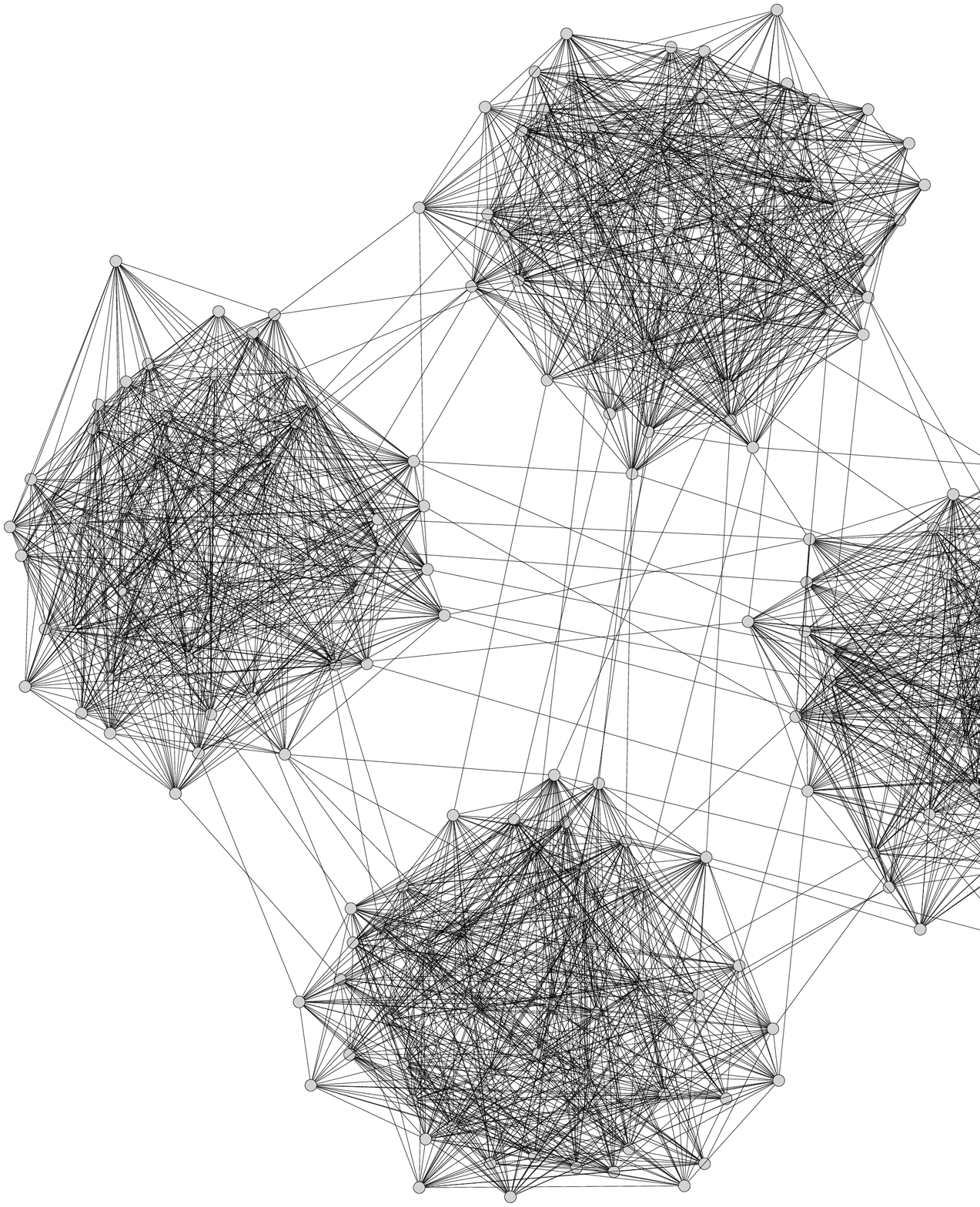}}
 \hspace{5mm}
   \subfigure[step $200$]{\includegraphics[width=0.48\linewidth]{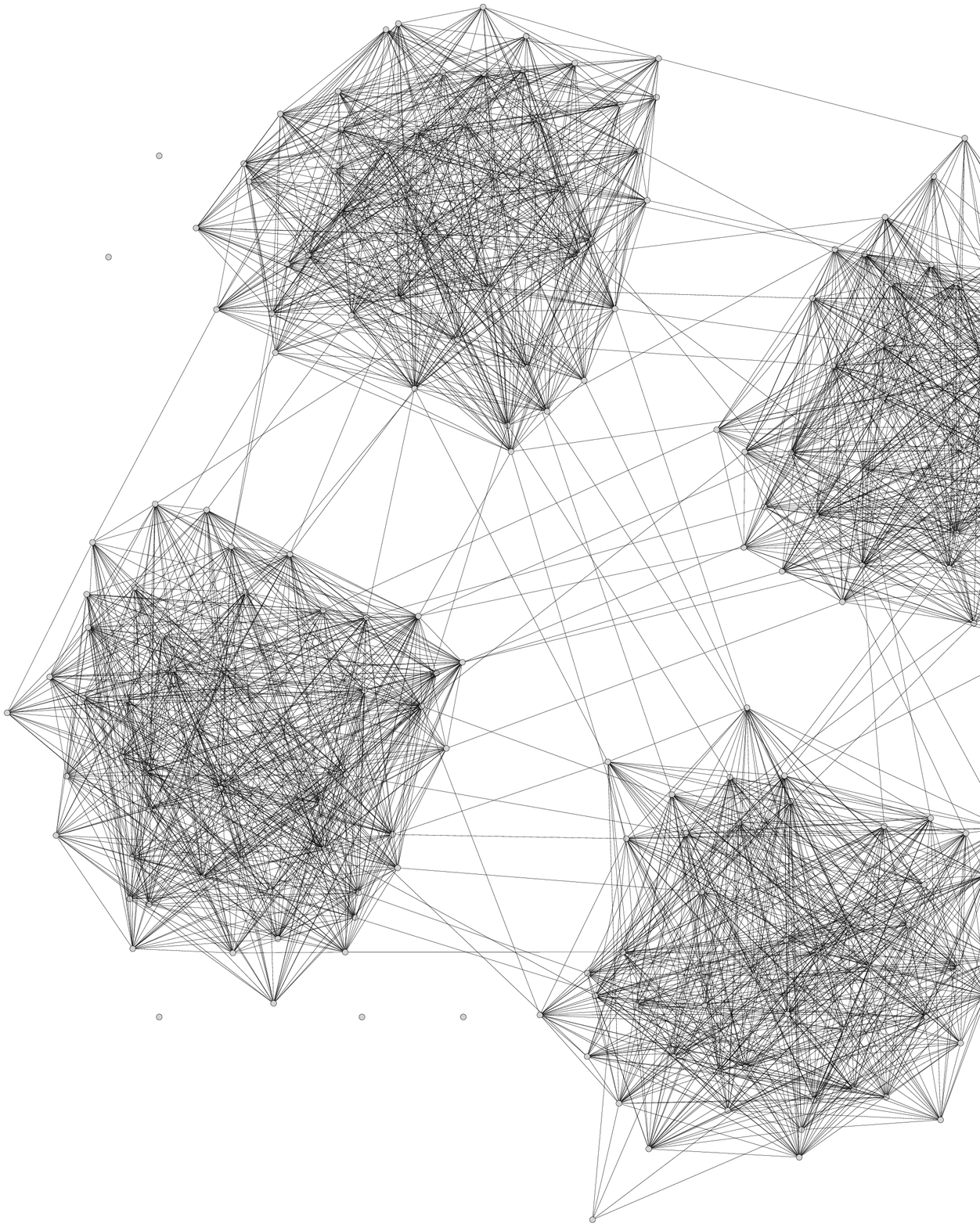}}
\caption{Two snapshots of the clustered network structure during evolution; node failure rate equal to the node join rate; $200$ nodes, $4$ clusters, $\gamma = 0.5$, $\omega = 0.2$ (hence on average there are $4 \omega = 10$ inter-cluster links)}
   \label{fig:structure_evolution}
\end{figure*}

First, it is interesting to observe that the ON mode does not alter the topology of the network. 
Just as an example, Figure \ref{fig:structure_evolution} reports a graphical representation of two snapshots of a network composed of $200$ nodes, at the beginning of the simulation and after $200$ simulation steps. In this case, a particularly dense network was considered.
It is possible to observe that the structure of the network is almost the same. Note that, due to the software employed to create these pictorial representations, the position of nodes in the network is not fixed, i.e.~a node that is in a certain position in the first snapshot is not positioned on the same coordinates in the other snapshot. Hence, only the general structure of the network should be considered here.
The nodes without links in the second picture are those currently inactive (failed).

\begin{figure}[thbp]
   \centering
   \includegraphics[width=\linewidth]{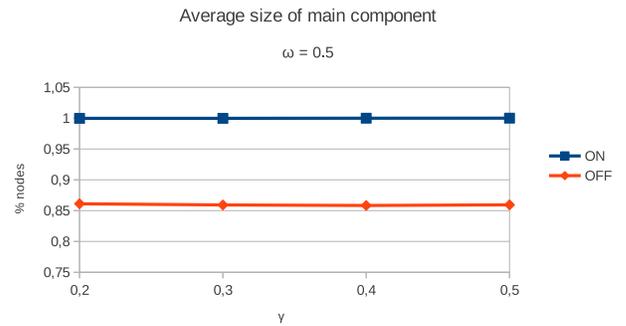}
   \caption{Average size of main component; $\gamma$ value varied -- clustered network; $200$ nodes, $4$ clusters (on average, there are $4\omega$ inter-cluster links)}
   \label{fig:comp_cluster}
\end{figure}

\begin{figure}[thbp]
   \centering
   \includegraphics[width=\linewidth]{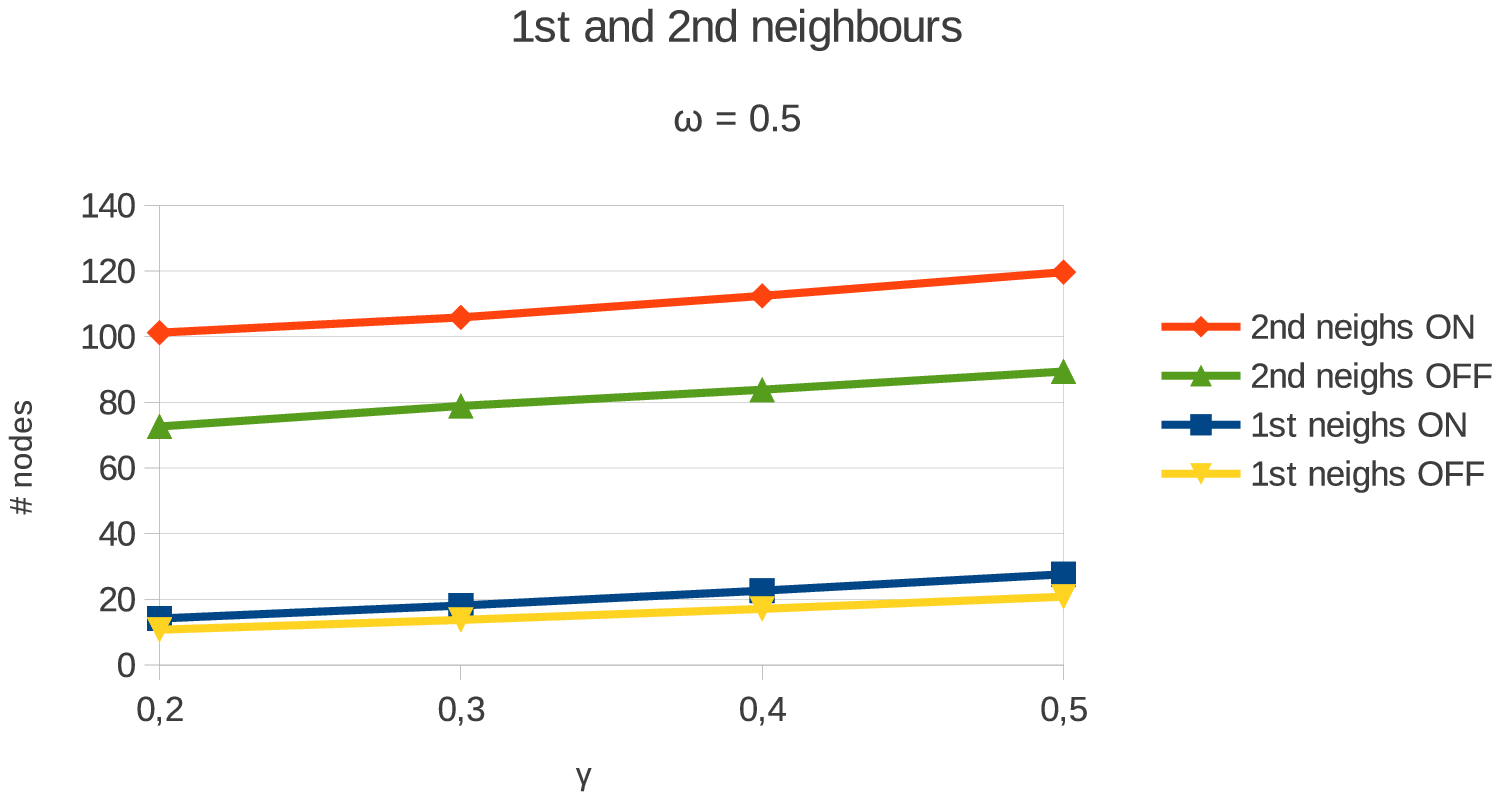}
   \caption{Average amount of neighbours; $\gamma$ value varied -- clustered network; $200$ nodes, $4$ clusters (on average, there are $4\omega$ inter-cluster links)}
   \label{fig:neighs_cluster}
\end{figure}

Figures \ref{fig:comp_cluster} and \ref{fig:neighs_cluster} show the average size of the main component and the average amount of 1st and 2nd neighbours, respectively, obtained in several configuration settings. In particular, in the two charts we vary on the x-axis the value of $\gamma$, i.e.~the probability of creating links among nodes of the same cluster, while leaving $\omega$ unaltered. 
In Figure \ref{fig:comp_cluster}, it is possible to appreciate that the ON mode guarantees that all nodes remain in the main component, while the OFF mode is not able to do it. However, $\gamma$ does not influence significantly the final result.
Figure \ref{fig:neighs_cluster} states that the ON mode guarantees a higher average amount of (1st and 2nd) neighbours, with respect to the OFF mode. The higher $\gamma$ the higher the amount of neighbours.

\begin{figure}[thbp]
   \centering
   \includegraphics[width=\linewidth]{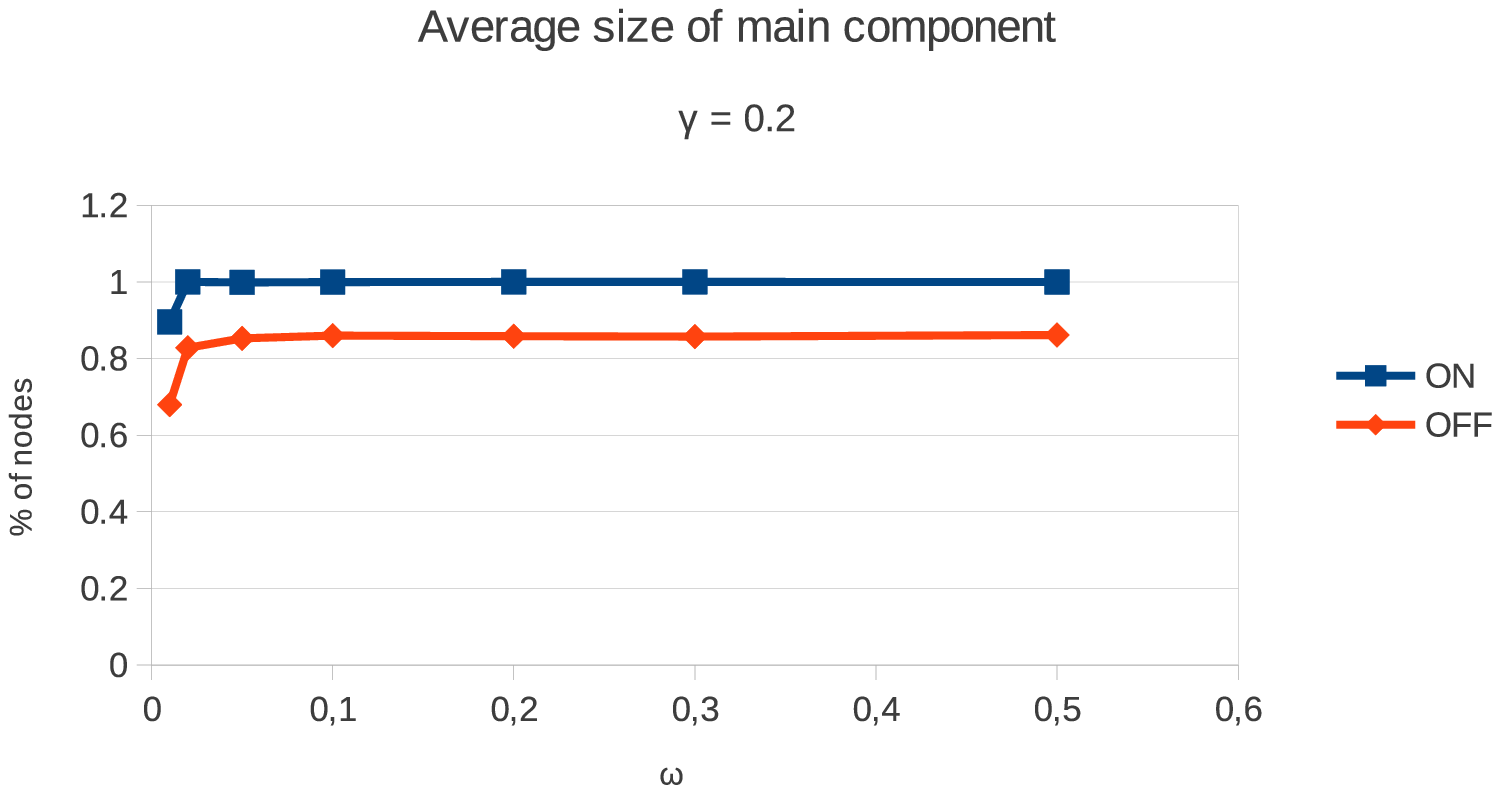}
   \caption{Average size of main component; $\omega$ value varied -- clustered network; $200$ nodes, $4$ clusters (on average, there are $4\omega$ inter-cluster links)}
   \label{fig:comp_omega}
\end{figure}

\begin{figure}[thbp]
   \centering
   \includegraphics[width=\linewidth]{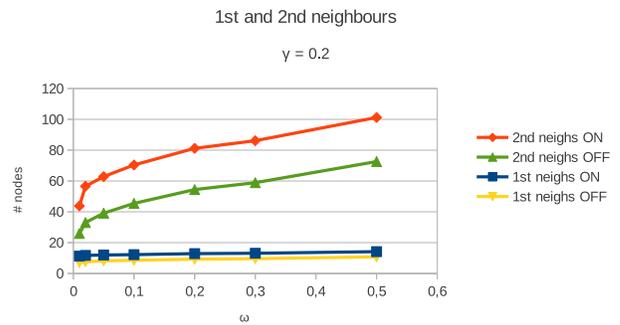}
   \caption{Average amount of neighbours; $\omega$ value varied -- clustered network; $200$ nodes, $4$ clusters (on average, there are $4\omega$ inter-cluster links)}
   \label{fig:neighs_omega}
\end{figure}

Figure \ref{fig:comp_omega} shows the average size of the main component in clustered networks, obtained when varying $\omega$. It is possible to observe that the size of the main component grows with $\omega$. In the ON mode, all nodes are in the main component, while a non-negligible portion of nodes remains outside the main component in the OFF mode. Figure \ref{fig:neighs_omega} shows the amount of 1st and 2nd neighbours; the ON mode provides a higher amount of neighbours with respect to the OFF mode. 
Moreover, the higher $\omega$ the higher the amount of 2nd neighbours and the (slightly) higher the amount of 1st neighbours. 

\subsection{Resilience to node faults}

It is interesting to understand how the protocol behaves when the network experiences several node faults. Thus, a simulation was performed where only failure events were generated. In particular, the network started with a certain topology; then nodes progressively fail until no nodes remained active. 
Of course, this is not a realistic scenario. However, this represents a worst case to assess resilience to faults.
Hence, it allows to understand if the failure management procedure explained in the previous section is able to maintain the network connected, despite the links removals due to node faults.

\begin{figure}[thbp]
   \centering
   \includegraphics[width=\linewidth]{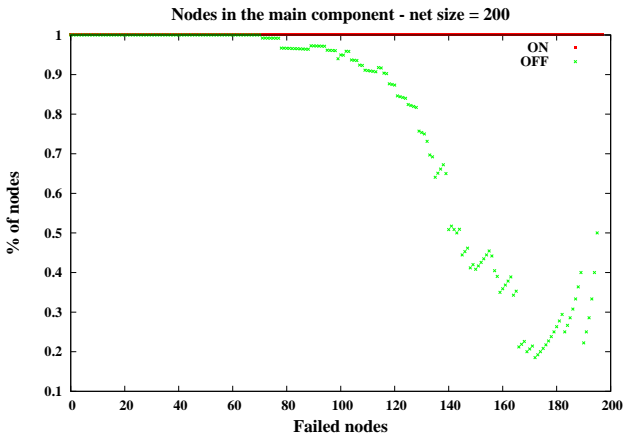}
   \caption{Main component size during a simulation; progressive node failures -- uniform network}
   \label{fig:fixed_comp}
\end{figure}

\begin{figure}[thbp]
   \centering
   \includegraphics[width=\linewidth]{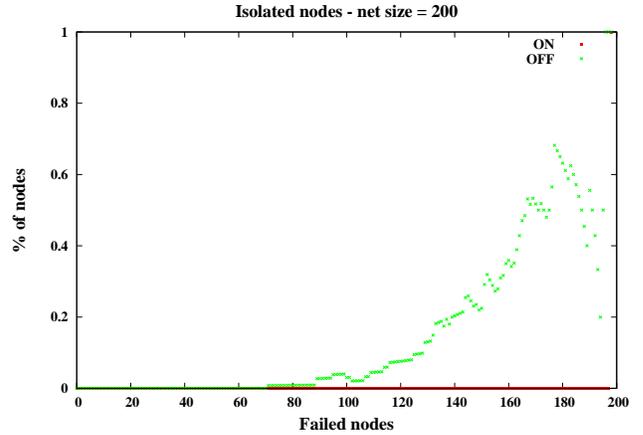}
   \caption{Amount of isolated nodes during a simulation; progressive node failures -- uniform network}
   \label{fig:fixed_isol}
\end{figure}

Figure \ref{fig:fixed_comp} shows the amount of nodes that remain in the main component during the evolution, while nodes continuously fail. In this case, we consider uniform networks composed of $200$ nodes, starting with a fixed degree equal to $5$. This figure shows the evolution of a specific network. We repeated the same experiment multiple times, varying the network size, the initial nodes' degree, and the seed for random generation, obtaining comparable results. 
It is possible to see that, in the OFF mode, at a certain point of the simulation the network gets disconnected and the percentage of active nodes in the main components decreases (at the end, the percentage of nodes in the main component might increase, meaning that isolated nodes have failed). Conversely, in the ON mode, active nodes remain connected in the same, single component. 
This is confirmed by looking at Figure \ref{fig:fixed_isol}, which shows the amount of nodes that remain isolated. While the percentage of isolated nodes increases in the OFF mode, no nodes remain isolated in the ON mode. It is worth adding that during the simulation of the OFF mode, we observed that some nodes formed some small components, which are not shown in the two figures.

\begin{figure}[thbp]
   \centering
   \includegraphics[width=\linewidth]{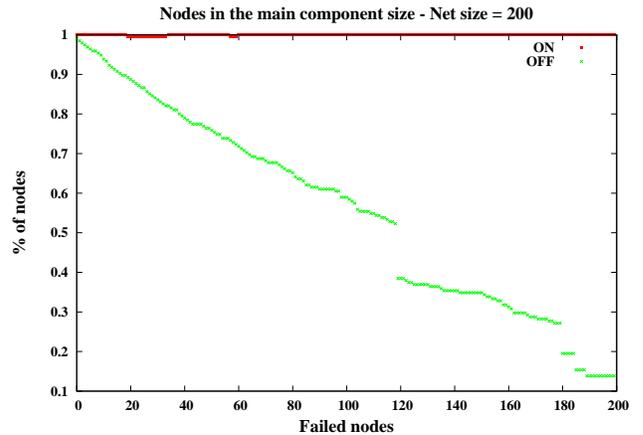}
   \caption{Main component size during a simulation; progressive node failures -- clustered network; $\gamma = 0.1, \omega = 0.1$; $200$ nodes, $4$ clusters (on average, there are $4\omega$ inter-cluster links)}
   \label{fig:cluster_comp}
\end{figure}

\begin{figure}[thbp]
   \centering
   \includegraphics[width=\linewidth]{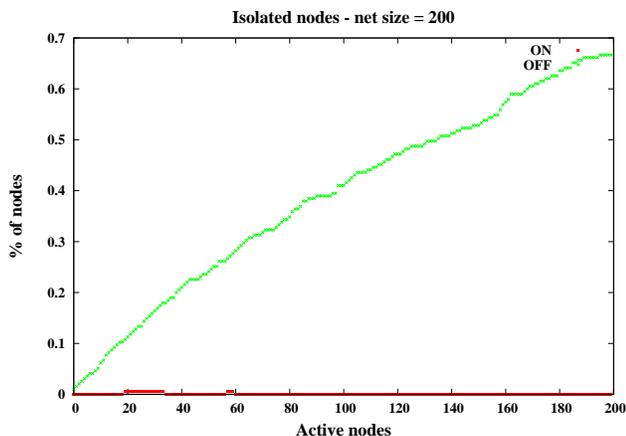}
\caption{Amount of isolated nodes during a simulation; progressive node failures -- clustered network; $\gamma = 0.1, \omega = 0.1$; $200$ nodes, $4$ clusters (on average, there are $4\omega$ inter-cluster links)}
   \label{fig:cluster_isol}
\end{figure}

Figures \ref{fig:cluster_comp} -- \ref{fig:cluster_isol} show the amount of nodes in the main component size and the amount of isolated nodes, respectively, when we simulate progressive node failures in clustered networks (they represent the results for a single simulation, which are in perfect accordance with all other runs for the same scenario). We show results for clustered networks of $200$ nodes, with $\gamma = 0.1, \omega = 0.1$. 
From these results, it is clear that (random) node faults have a strong impact on network connectivity. In fact, in the OFF mode the network gets disconnected and the number of active nodes in the main component progressively decreases with node faults. 
During the simulation we noticed the formation of minor components (not shown in these figures). Moreover, an increasing amount of active nodes gets isolated from the rest of the network, as shown in Figure \ref{fig:cluster_isol}. 
Conversely, the ON mode allows the network to remain connected till the end of the simulation, i.e.~until all nodes failed. This is an important result that confirms that the use of a simple local strategy guarantees the resilience of the unstructured network, whatever its topology.

\section{Conclusions}\label{sec:conc}

Outcomes confirm that it is possible to guarantee resilience to node failures in unstructured P2P overlay networks. The use of simple and local protocols avoid network disconnections. 
To demonstrate this, a very simple approach has been studied which requires knowledge at nodes of 1st and 2nd neighbours. The protocol avoids that a too high amount of links are created among nodes, to react to a single node departure, thus preserving the network topology.
In essence, the presented approach guarantees that the overlay network remains connected, without the need (and the related costs) to add a structure to the overlay.

\def\newblock{}

\end{document}